\documentclass{article} 
\usepackage{iclr2017_workshop,times}
\usepackage{hyperref}
\usepackage{url}

\usepackage{amsmath}
\usepackage{amsthm}
\usepackage{amsfonts}

\usepackage{graphicx}

\usepackage{wrapfig}

\title{ Audio Super-Resolution using Neural Nets }

\author{Volodymyr Kuleshov, S. Zayd Enam, and Stefano Ermon \\
Department of Computer Science, \\
Stanford University\\
\texttt{\{kuleshov,ermon\}@cs.stanford.edu}\\
\texttt{zayd@stanford.edu}
}

\begin{document}

\maketitle

\begin{abstract}
We introduce a new audio processing technique that increases the sampling rate of  signals such as speech or music using deep convolutional neural networks.
Our model is trained on pairs of low and high-quality audio examples; at test-time, it predicts missing samples within a low-resolution signal in an interpolation process 
similar to image super-resolution.
Our method is simple and does not involve specialized audio processing techniques; in our experiments, it outperforms baselines 
on standard speech and music benchmarks at upscaling ratios of $2\times$, $4\times$, and $6\times$. The method has practical applications in telephony, compression, and text-to-speech generation; it demonstrates the effectiveness of convolutional architectures on an audio generation task.

\end{abstract}

\section{Introduction}

The generative modeling of audio signals is a fundamental problem at the intersection of signal processing and machine learning; recent learning-based algorithms have enabled advances in speech recognition~\citep{hinton2012deep}, audio synthesis \citep{DBLP:journals/corr/OordDZSVGKSK16, mehri2016samplernn}, music recommendation systems~\citep{coviello2012multivariate, wang2014improving, liang2015content}, and in many other areas~\citep{acevedo2009automated}. Audio processing also raises basic research questions pertaining to time series and generative modeling~\citep{haykin2005cocktail,bilmes2004graphical}.

One of the most significant recent advances in machine learning-based audio processing has been the ability to directly model {\em raw signals} in the time domain using neural networks \citep{DBLP:journals/corr/OordDZSVGKSK16, mehri2016samplernn}. Although this affords us the maximum modeling flexibility, it is also computationally expensive, requiring us to handle $>10,000$ audio samples at every second. 

In this paper, we explore new lightweight modeling algorithms for audio.
In particular, we focus on a specific audio generation problem called {\em bandwidth extension}, in which the task is to reconstruct high-quality audio from a low-quality, down-sampled input containing only a small fraction (15-50\%) of the original samples. We introduce a new neural network-based technique for this problem that is inspired image super-resolution algorithms~\citep{Dong:2016:ISU:2914182.2914303}, which use machine learning techniques to interpolate a low-resolution image into a higher-resolution one. Learning-based methods often perform better in this context than general-purpose interpolation schemes such as splines because they leverage sophisticated domain-specific models of the appearance of natural signals.
 
As in image super-resolution, our model is trained on pairs of low and high-quality samples; at test-time, it predicts the missing samples of a low-resolution input signal. Unlike recent neural networks for generating raw audio, our model is fully feedforward and can be run in real-time. In addition to having multiple practical applications, our method also suggests new ways to improve existing generative models of audio.

\subsection{Contributions}


From a practical perspective, our technique has applications in telephony, compression, text-to-speech generation, forensic analysis, and in other domains. It outperforms baselines at $2\times$, $4\times$, and $6\times$ upscaling ratios, while also being significantly simpler than previous methods. Whereas most existing audio enhancement methods make substantial use of signal processing theory, our approach is conceptually very simple and requires no specialized knowledge to implement. Our neural networks are simply trained to map one audio time series into another. Our approach is also among the first to use convolutional architectures for bandwidth extension; as a result, it scales better with dataset size and computational resources relative to current alternatives.

From a generative modeling perspective, our work demonstrates that purely feedforward architectures operating in a non-discretized output space can achieve good performance on an important audio generation task. 
This hints at the possibility of designing improved generative models for audio that combine both feedforward and recurrent components.

\section{Setup and background}

\paragraph{Audio processing.} 
We represent an audio signal as a function $s(t) : [0,T] \to \mathbb{R}$, where $T$ is the duration of the signal (in seconds) and $s(t)$ is the amplitude at $t$. Taking a digital measurement of $s$ requires us to discretize the continuous function $s(t)$ into a vector $x(t) :  \{\frac{1}{R}, \frac{2}{R},..., \frac{RT}{R}\} \to \mathbb{R}$. We refer to $R$ as the {\em sampling rate} of $x$ (in Hz). Sampling rates may range from 4 KHz (low-quality telephone speech) to 44 Khz (high-fidelity music).

In this work, we interpret $R$ as the resolution of $x$; our goal is to increase the resolution of audio samples by predicting $x$ from a fraction of its samples taken at $\{\frac{1}{R}, \frac{2}{R},..., \frac{RT}{R}\}$. Note that by basic signal processing theory, this is equivalent to predicting the higher frequencies of $x$.

\paragraph{Bandwidth extension.}

Audio upsampling has been studied in the audio processing community under the name {\em bandwidth extension} \citep{ekstrand2002bandwidth,larsen2005audio}. Several learning-based approaches have been proposed, including Gaussian mixture models \citep{cheng1994statistical, park2000narrowband} and neural networks \citep{li2015dnn}. These methods typically involve hand-crafted features and use relatively simple models (e.g., neural networks with at most 2-3 densely connected layers) that are often part of a larger, more complex systems. 
In comparison, our method is conceptually simple (operating directly on the raw audio signal), scalable (our neural networks are fully convolutional and fully feed-forward), more accurate, and is also among the few to have been tested on non-speech audio.

%


\section{Method}

\begin{figure}[t]
\vspace{-7mm}
\begin{center}
\includegraphics[width=12cm]{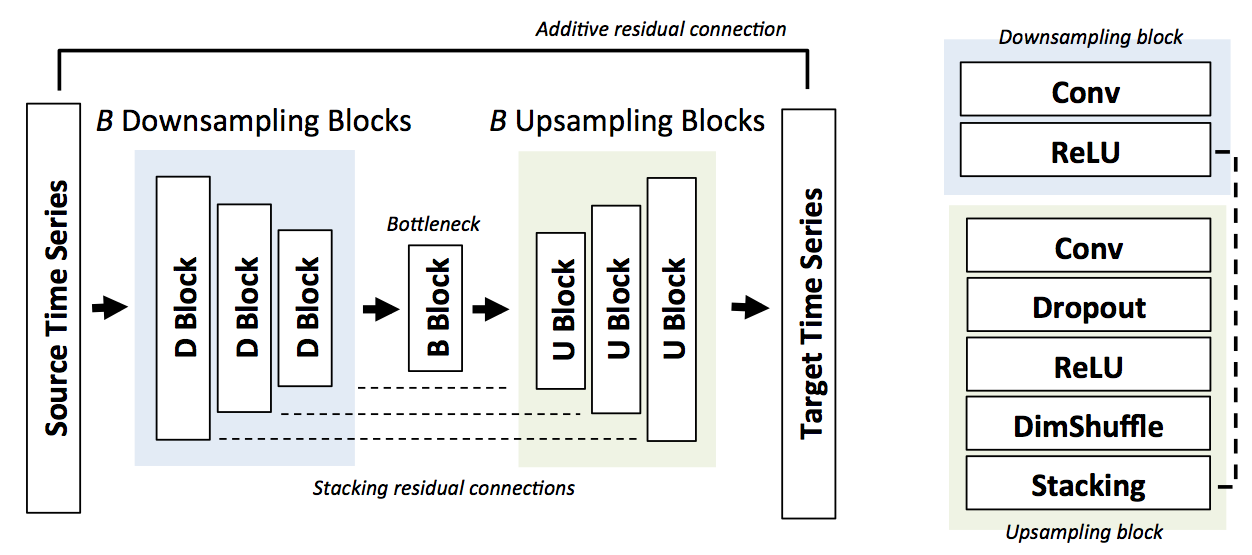}
\end{center}
\caption{Deep residual network used for audio super-resolution. We extract features via $B$ residual blocks; upscaling is done via stacked SubPixel layers.}
\end{figure}

\subsection{Setup}

Given a low resolution signal $x = \{x_{1/R_1}, ... x_{R_1 T_1 / R_1}\}$ sampled at a rate $R_1$, our goal is to reconstruct a high-resolution version $y = \{y_{1 / R_2}, ... y_{R_2 T_2 / R_2}\}$ of $x$ that has a sampling rate $R_2 > R_1$. For example, $x$ may be a voice signal transmitted via a standard telephone connection at 4 KHz; $y$ may be a high-resolution 16 KHz reconstruction of the orignal.
We use $r = R_2/R_1$ to denote the {\em upsampling ratio} of the two signals, which in our work equals $r=2,4,6$. We thus expect that $y_{r t / R_2} \approx x_{t / R_1}$ for $t = 1,2,...,T_1 R_1$.

To recover the under-defined signal, we learn a model $p(y \vert x)$ of the higher-resolution $y$, conditioned on its low-resolution instantiation $x$. 
We assume that the relationship between the time series $x, y$ follows the equation
$y = f_\theta(x) + \epsilon,$
where $\epsilon \sim \mathcal{N}(0,1)$ is Gaussian noise and $f_\theta$ is a model parametrized by $\theta$. Our framework also extends to more complex noise models which the user may provide as a prior or that may be themselves parametrized by the model (similarly to how one parametrizes the normal distribution in a variational autoencoder).

The above formulation naturally leads to a mean squared error (MSE) objective  
\begin{equation}
\ell(\mathcal{D}) = \frac{1}{n} \sqrt{ \sum_{i=1}^n ||y_i - f_\theta(x_i)||_2^2 }
\end{equation}
for determining the parameters $\theta$ based on a dataset $\mathcal{D} = \{x_i, y_i\}_{i=1}^n$ of source/target time series pairs. Since our model is fully convolutional, we may take the $x_i, y_i$ to be small patches sampled from the full time series.

\begin{figure}[t]
\begin{center}
\includegraphics[width=14cm]{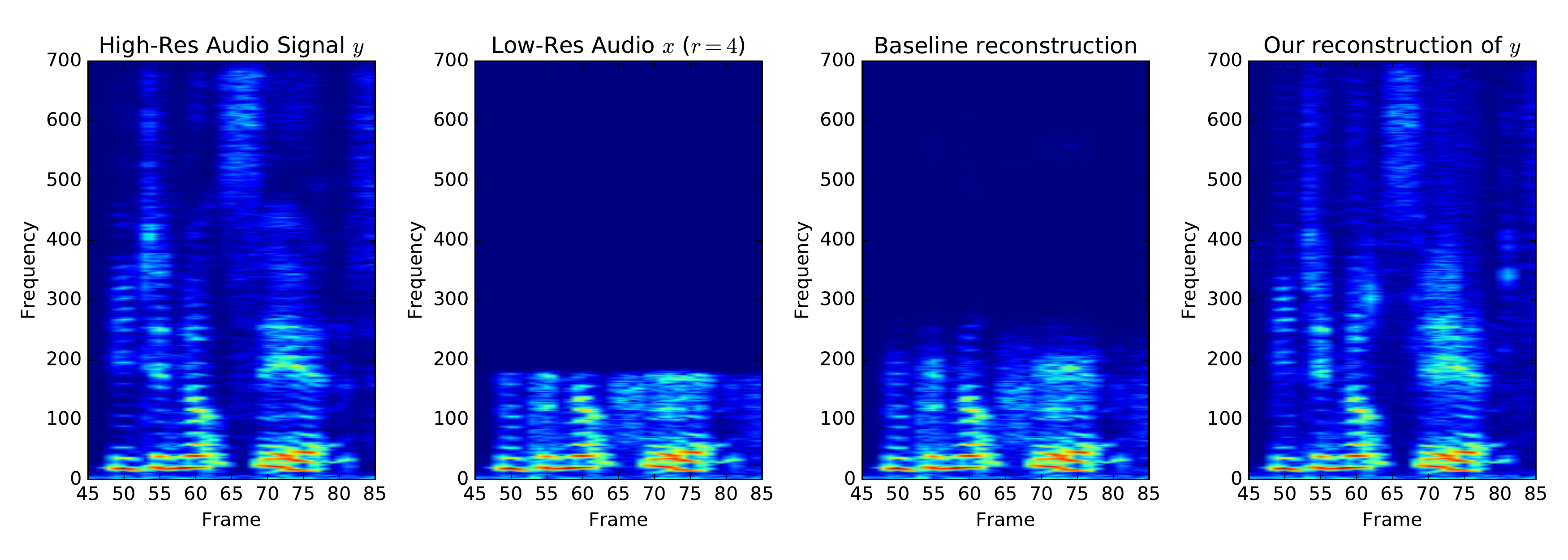}
\end{center}
\vspace{-5mm}
\caption{Audio super-resolution visualized using spectrograms. A high-quality speech signal (leftmost) is subsampled at $r=4$, resulting in the loss of high frequencies (2nd from left). We recover the missing signal using a trained neural network (rightmost), greatly outperforming the cubic baseline (second from right). }
\end{figure}

\subsection{Model Architecture} 

We parametrize the function $f$ with a deep convolutional neural network with residual connections; our neural network architecture is based on ideas from \citet{ShiCHTABRW16}, \citet{Dong:2016:ISU:2914182.2914303}, and \citet{pix2pix2016}, and is shown in Figure 1.
We highlight its main features below.


\paragraph{Bottleneck architecture.}

Our model contains $B$ successive downsampling and upsampling {\em blocks}: each performs a convolution, batch normalization, and applies a ReLU non-linearity. Downsampling block $b=1,2,...,B$ contains $\max(2^{6+b}, 512)$ convolutional filters of length $\min(2^{7-b}+1, 9)$ and a stride of $2$. Upsampling block $b$ has $\max(2^{7+(B-b+1)}, 512)$ filters of length $\min(2^{7-(B-b+1)}+1, 9)$.

Thus, at a downsampling step, we halve the spatial dimension and double the filter size; during upsampling, this is reversed. This bottleneck architecture is inspired by auto-encoders, and is known to encourage the model to learn a hierarchy of features. For example, on an audio task, bottom layers may extract wavelet-style features, while higher ones may correspond to phonemes \cite{DBLP:conf/nips/AytarVT16}. Note that the model is fully convolutional, and may run on input sequences of arbitrary length.

\paragraph{Skip connections.}

When the source series $x$ is similar to the target $y$, downsampling features will be also be useful for upsampling \citep{pix2pix2016}. We thus add additional skip connections which stack the tensor of $b$-th downsampling features with the $(B-b+1)$-th tensor of upsampling features. We also add an additive residual connection from the input to the final output: the model thus only needs to learn $y-x$, which in practice speeds up training.

\paragraph{Subpixel shuffling layer.}

In order to increase the time dimension during upscaling, we have implemented a one-dimensional version of the Subpixel layer of \citet{ShiCHTABRW16}, which has been shown to be less prone to produce artifacts \citep{odena2016deconvolution}.

An upscaling block's convolution maps an input tensor of dimension $F \times d$ into one of size $F/2 \times d$.
The subpixel layer reshuffles this $F/2 \times d$ tensor into another one of size $F/4 \times 2d$ (while preserving the tensor entries intact); these are concatenated with $F/4$ features from the downsampling stage, for a final output of size $F/2 \times 2d$. Thus, we have halved the number of filters and doubled the spatial dimension.

\section{Experiments}

\paragraph{Datasets.} 

We use the VCTK dataset \citep{yamagishienglish} --- which contains 44 hours of data from 108 different speakers --- and the Piano dataset of \citet{mehri2016samplernn} (10 hours of  Beethoven sonatas). 
We generate low-resolution audio signal from the 16 KHz originals by applying an order 8 Chebyshev type I low-pass filter before subsampling the signal by the desired scaling ratio.

We evaluate our method in three regimes.
The {\sc SingleSpeaker} task trains the model on the first 223 recordings of VCTK Speaker 1 (about 30 mins) and tests on the last 8 recordings.
The {\sc MultiSpeaker} task assesses our ability to generalize to new speakers. We train on the first 99 VCTK speakers and test on the 8 remaining ones; our recordings feature different voices and accents (Scottish, Indian, etc.)
Lastly, the {\sc Piano} task extends audio-super resolution to non-vocal data; we use the standard 88\%-6\%-6\% data split.

\paragraph{Methods.}

We compare our method relative to two baselines: a cubic B-spline --- which corresponds to the bicubic upsampling baseline used in image super-resolution --- and the recent neural network-based technique of \citet{li2015dnn}, 


The latter approach takes as input the short-time Fourier transform (STFT) of the input and predicts directly the phase and the magnitudes of the high frequency components using a dense neural network with three hidden layers of size 2048 and ReLU nonlinearities.
\citet{li2015dnn} have shown that this method is preferred over Gaussian Mixture Models in 84\% of cases in a user study.
This model requires that the scaling ratio be a power of $2$, hence it is not applicable when $r=6$.

We instantiate our model with $B=4$ blocks and train it for 400 epochs on patches of length 6000 (in the high-resolution space) using the ADAM optimizer with a learning rate of $10^{-4}$. To ensure source/target series are of the same length, the source input is pre-processed with cubic upscaling.
We do not compare against previously-proposed matrix factorization techniques \citep{Bansal05bandwidthexpansion,DBLP:conf/ismir/LiangHE13}, as they are typically trained on $<$ 10 input examples \citep{DBLP:conf/mlsp/SunM13} (due to the cost of jointly factorizing a large number of matrices), and do not scale to the size of our datasets.


\paragraph{Metrics}

Given a reference signal $y$ and an approximation $x$, the Signal to Noise Ratio (SNR) is defined as
\begin{equation}
\text{SNR}(x,y) = 10 \log \frac{|| y ||_2^2}{|| x-y ||_2^2}.
\end{equation}
The SNR is a standard metric used in the signal processing literature.
The Log-spectral distance (LSD) \citep{gray1976distance} measures the reconstruction quality of individual frequencies as follows:
\begin{equation}
\text{LSD}(x,y) = \frac{1}{L} \sum_{\ell=1}^L \sqrt{\frac{1}{K} \sum_{k=1}^K \left( X(\ell, k) - \hat X(\ell, k) \right)^2 }, 
\end{equation}
where $X$ and $\hat X$ are the log-spectral power magnitudes of $y$ and $x$, respectively. These are defined as $ X = \log |S|^2 $, where $S$ is the short-time Fourier transform (STFT) of the signal. We use $\ell$ and $k$ index frames and frequencies, respectively; in our experiments, we used frames of length 2048.

\paragraph{Evaluation} 

\begin{wraptable}{r}{0.5\textwidth}
\vspace{-3mm}
\begin{tabular}{rcccc|c}
& \multicolumn{4}{c}{MultiSpeaker Sample} & \\
\hline
 & 1 & 2 & 3 & 4 & Average \\
\hline
Ours & 69 & 75 & 64 & 37 & 61.3 \\
DNN & 51 & 55 & 66 &  53 & 56.3 \\
Spline & 31 & 25 & 38 & 47 & 35.3 \\
\hline
\end{tabular}
\caption{MUSHRA user study scores. We show scores for each sample, averaged individual users. Average across all samples is also displayed}
\label{mos}
\end{wraptable}

The results of our experiments are summarized in Table 2. Our objective metrics show an improvement of 1-5 dB over the baselines, with the strongest improvements at higher upscaling factors.
Although, the spline baseline achieves a high SNR, its signal often lacks higher frequencies; the LSD metric is better at identifying this problem. 
Our technique also improves over the DNN baseline; our convolutional architecture appears to use our  modeling capacity more efficiently than a dense neural network, and we expect such architectures will soon be more widely used in audio generation tasks.





\begin{table}[t]
\label{multispeaker}
\begin{center}
\begin{tabular}{lr|rrr|rrr|rrr}
& & \multicolumn{3}{c}{SingleSpeaker}  & \multicolumn{3}{c}{MultiSpeaker} & \multicolumn{3}{c}{Piano}                  \\
\hline
Ratio & Obj. & Spline & DNN & Ours  & Spline & DNN & Ours  & Spline & DNN & Ours \\
\hline
$r=2$    & SNR & 20.3 & 20.1 & 21.1 & 19.7 & 19.9 & 20.7 & 29.4 & 29.3 & 30.1 \\
    & LSD & 4.5 & 3,7 & 3.2 & 4.4 & 3.6 & 3.1 & 3.5 & 3.4 & 3.4 \\
$r=4$    & SNR & 14.8 & 15.9 & 17.1  & 13.0 & 14.9 & 16.1  & 22.2 & 23.0 & 23.5 \\
    & LSD & 8.2 & 4.9 & 3.6  & 8.0 & 5.8 & 3.5 & 5.8 & 5.2 & 3.6 \\
$r=6$    & SNR & 10.4 & n/a & 14.4  & 9.1 & n/a & 10.0 & 15.4 & n/a & 16.1 \\
    & LSD & 10.3 & n/a & 3.4  & 10.1 & n/a & 3.7 & 7.3 & n/a & 4.4 \\
\end{tabular}
\end{center}
\caption{Accuracy evaluation of audio-super resolution methods (in dB) on each of the three super-resolution tasks at upscaling ratios $r=2,4,6$.}
\end{table}

Next, we confirmed our objective experiments with a study in which human raters were asked to assess the quality of super-resolution using a MUSHRA (MUltiple Stimuli with Hidden Reference and Anchor) test. For each trial an audio sample was upscaled using different techniques\footnote{We have posted a our set of samples to: {\texttt https://kuleshov.github.io/audio-super-res/}.}. We collected four VCTK speaker recordings audio samples from the {\sc MultiSpeaker} testing set. For each recording, we collected the original utterance, a downsampled version at $r=4$, as well as signals super-resolved using Splines, DNNs, and our model (six versions in total). We recruited 10 subjects and used an online survey to ask each of them to rate each sample on a scale of 0 (extremely bad) to 100 (excellent) reconstruction.
The results from the experiment are summarized in Table \ref{mos}.
Our method ranked as being the best out of the three upscaling techniques.

\paragraph{Domain adaptation.}

\begin{wraptable}{r}{0.45\textwidth}
\vspace{-3mm}
\begin{small}
\begin{tabular}{rrrrr}
\hline
 & \multicolumn{2}{c}{LPF {\scriptsize (Test)}} & \multicolumn{2}{c}{No LPF {\scriptsize (Test)}} \\
  & SNR & LSD & SNR & LSD  \\
\hline
LPF {\scriptsize (Train)}  & 30.1 & 3.4 & 0.42 & 4.5   \\
No LPF {\scriptsize (Train)}  & 0.43 & 4.4 & 33.2 & 3.3  \\
\hline
\end{tabular}
\caption{Sensitivity of the model to whether low-resolution audio was subject to a low-pass filter (LPF) in dB.
}\label{scaling}
\vspace{-1mm}
\end{small}
\end{wraptable}

We tested the sensitivity of our method to out-of-distribution input via an audio super-resolution experiment in which the training set did not use a low-pass filter, while the test set did, and vice-versa.
We focused on the {\sc Piano} task and $r=2$. The output from the model was noisier than expected, indicating that generalization is an important practical concern. 
We suspect this behavior may be common in super-resolution algorithms, but has not been widely documented. A potential solution would be to train on data that has been generated using multiple techniques.

In addition, we examined the ability of our model to generalize from speech to music and vice versa. 
We found that switching domains produced noisy output, again highlighting the specialization of the model.


\paragraph{Architectural analysis.}

We examined the importance of our various architectural design choices via an ablation analysis on the {\sc MultiSpeaker} audio super-resolution task using an upscaling ratio of $r=4$. The adjacent figure displays the result: the green-ish line display the validation set $\ell_2$ loss of the original model over time; the yellow curve removes the additive residual connection; the green curve further removes the additive skip connection (while preserving the same total number of filters). This shows that symmetric skip connections are crucial for attaining good performance; additive connections add an additional small, but perceptible, improvement.

\begin{wrapfigure}{r}{0.53\textwidth}
\label{fig:curves}
\vspace{-5mm}
\includegraphics[width=8cm]{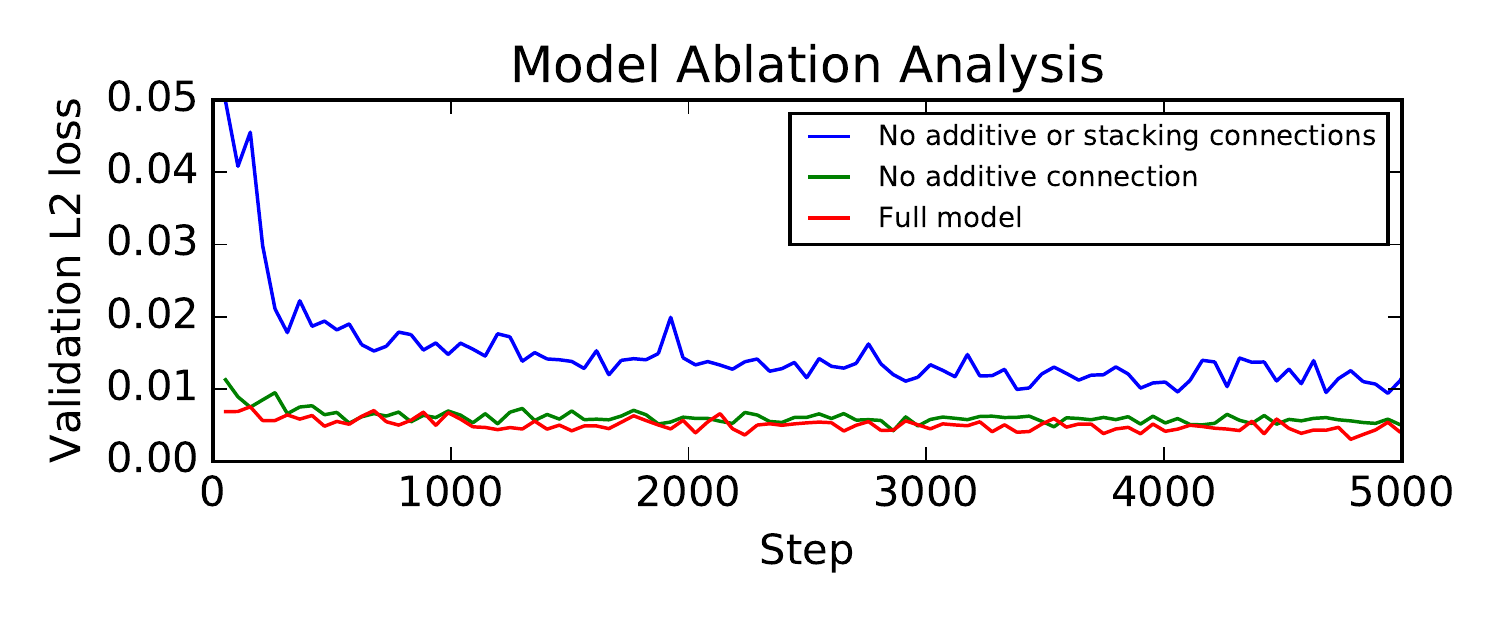}
\vspace{-7mm}
\begin{small}
\caption{Model ablation analysis on the MultiSpeaker audio super-resolution task with $r=4$.
}\end{small}
\end{wrapfigure}



\paragraph{Computational performance.}

Our model is computationally efficient and can be run in real time. 
On the {\sc Piano} task (where all input signals are 12s in length), our method processed a single second of audio in 0.11s on average on a Titan X GPU. Training our models, however, required about 2 days for the {\sc MultiSpeaker} task.
Unlike sequence-to-sequence architectures our model does not require the complete input sequence in order to begin generating an output sequence.

\subsection{Limitations}

Finally, to explore the limits of our approach, we evaluated our method on the MagnaTagATune dataset, which consists of about 200 hours of music from 188 different genres. This dataset is larger and much more diverse that the ones we considered so far. We found that our model underfit the dataset, with very little reduction in the training error, and no improvement over the spline baseline. Other learning-based baselines fared similarly. However, we expect improved results with a larger model and more computational resources.

\section{Previous Work and Discussion}

\paragraph{Time series modeling.}

In the machine learning literature, time series signals have most often been modeled with auto-regressive models, of which variants of recurrent networks are a special case \citep{gers2001applying,45168,mehri2016samplernn}.
Our approach instead generalizes conditional modeling ideas used in computer vision for tasks such as
image super-resolution \citep{Dong:2016:ISU:2914182.2914303,DBLP:journals/corr/LedigTHCATTWS16} or colorization \citep{zhang2016colorful}. 

We identify a broad class of conditional time series modeling problems that arise in signal processing, biomedicine, and other fields and that are characterized by a natural alignment among source/target series pairs and differences that are well-represented by local transformations. We propose a general architecture for such problems and show that it works well in different domains. 

\paragraph{Bandwidth extension.}

%
Existing learning-based approaches include Gaussian mixture models \citep{cheng1994statistical,park2000narrowband,pulakka2011speech}, linear predictive coding \citep{bradbury2000linear}, and neural networks \citep{li2015dnn}. 
Our work proposes the first convolutional architecture, which we find to scale better with dataset size and outperform recent, specialized methods.
Moreover, while existing techniques involve many hand-crafted features (see e.g., \citet{pulakka2011speech}); our approach is fully domain-agnostic.


%
%

\paragraph{Audio applications.} In telephony, commercial efforts are underway to transmit voice at higher rates (typically 16 Khz) in specific handsets; audio-super resolution is a step towards recreating this experience in software. Similar applications could be found in compression, text-to-speech generation, and forensic analysis.
More generally, our work demonstrates the effectiveness of feedforward convolutional architectures on an audio generation task.

\section{Conclusion}

Machine learning techniques based on deep neural networks have been successful at solving under-defined problems in signal processing such as image super-resolution, colorization, in-painting, and many others. 
Learning-based methods often perform better in this context than general-purpose algorithms because they leverage sophisticated domain-specific models of the appearance of natural signals.

In this work, we proposed new techniques that use this insight to upsample audio signals. Our technique extends previous work on image super-resolution to the audio domain; it outperforms previous bandwidth extension approaches on both speech and non-vocal music. Our approach is fast and simple to implement, and has applications in telephony, compression, and text-to-speech generation. It also demonstrates the effectiveness of feedforward architectures on an important audio generation task, suggesting new directions for generative audio modeling.


%

\bibliography{all}

\begin{thebibliography}{30}
\providecommand{\natexlab}[1]{#1}
\providecommand{\url}[1]{\texttt{#1}}
\expandafter\ifx\csname urlstyle\endcsname\relax
  \providecommand{\doi}[1]{doi: #1}\else
  \providecommand{\doi}{doi: \begingroup \urlstyle{rm}\Url}\fi

\bibitem[Acevedo et~al.(2009)Acevedo, Corrada-Bravo, Corrada-Bravo,
  Villanueva-Rivera, and Aide]{acevedo2009automated}
Miguel~A Acevedo, Carlos~J Corrada-Bravo, H{\'e}ctor Corrada-Bravo, Luis~J
  Villanueva-Rivera, and T~Mitchell Aide.
\newblock Automated classification of bird and amphibian calls using machine
  learning: A comparison of methods.
\newblock \emph{Ecological Informatics}, 4\penalty0 (4):\penalty0 206--214,
  2009.

\bibitem[Aytar et~al.(2016)Aytar, Vondrick, and
  Torralba]{DBLP:conf/nips/AytarVT16}
Yusuf Aytar, Carl Vondrick, and Antonio Torralba.
\newblock Soundnet: Learning sound representations from unlabeled video.
\newblock In \emph{Advances in Neural Information Processing Systems 29: Annual
  Conference on Neural Information Processing Systems 2016, December 5-10,
  2016, Barcelona, Spain}, pp.\  892--900, 2016.
\newblock URL
  \url{http://papers.nips.cc/paper/6146-soundnet-learning-sound-representations-from-unlabeled-video}.

\bibitem[Bansal et~al.(2005)Bansal, Raj, and
  Smaragdis]{Bansal05bandwidthexpansion}
Dhananjay Bansal, Bhiksha Raj, and Paris Smaragdis.
\newblock Bandwidth expansion of narrowband speech using non-negative matrix
  factorization.
\newblock In \emph{in Proc. Interspeech}, 2005.

\bibitem[Bilmes(2004)]{bilmes2004graphical}
Jeffrey~A Bilmes.
\newblock Graphical models and automatic speech recognition.
\newblock In \emph{Mathematical foundations of speech and language processing},
  pp.\  191--245. Springer, 2004.

\bibitem[Bradbury(2000)]{bradbury2000linear}
Jeremy Bradbury.
\newblock Linear predictive coding.
\newblock \emph{Mc G. Hill}, 2000.

\bibitem[Cheng et~al.(1994)Cheng, O'Shaughnessy, and
  Mermelstein]{cheng1994statistical}
Yan~Ming Cheng, Douglas O'Shaughnessy, and Paul Mermelstein.
\newblock Statistical recovery of wideband speech from narrowband speech.
\newblock \emph{IEEE Transactions on Speech and Audio Processing}, 2\penalty0
  (4):\penalty0 544--548, 1994.

\bibitem[Coviello et~al.(2012)Coviello, Vaizman, Chan, and
  Lanckriet]{coviello2012multivariate}
Emanuele Coviello, Yonatan Vaizman, Antoni~B Chan, and Gert~RG Lanckriet.
\newblock Multivariate autoregressive mixture models for music auto-tagging.
\newblock In \emph{ISMIR}, pp.\  547--552, 2012.

\bibitem[Dong et~al.(2016)Dong, Loy, He, and
  Tang]{Dong:2016:ISU:2914182.2914303}
Chao Dong, Chen~Change Loy, Kaiming He, and Xiaoou Tang.
\newblock Image super-resolution using deep convolutional networks.
\newblock \emph{IEEE Trans. Pattern Anal. Mach. Intell.}, 38\penalty0
  (2):\penalty0 295--307, February 2016.
\newblock ISSN 0162-8828.
\newblock \doi{10.1109/TPAMI.2015.2439281}.
\newblock URL \url{http://dx.doi.org/10.1109/TPAMI.2015.2439281}.

\bibitem[Ekstrand(2002)]{ekstrand2002bandwidth}
Per Ekstrand.
\newblock Bandwidth extension of audio signals by spectral band replication.
\newblock In \emph{in Proceedings of the 1st IEEE Benelux Workshop on Model
  Based Processing and Coding of Audio (MPCA’02}. Citeseer, 2002.

\bibitem[Gers et~al.(2001)Gers, Eck, and Schmidhuber]{gers2001applying}
Felix~A Gers, Douglas Eck, and J{\"u}rgen Schmidhuber.
\newblock Applying lstm to time series predictable through time-window
  approaches.
\newblock In \emph{International Conference on Artificial Neural Networks},
  pp.\  669--676. Springer, 2001.

\bibitem[Gray \& Markel(1976)Gray and Markel]{gray1976distance}
Augustine Gray and John Markel.
\newblock Distance measures for speech processing.
\newblock \emph{IEEE Transactions on Acoustics, Speech, and Signal Processing},
  24\penalty0 (5):\penalty0 380--391, 1976.

\bibitem[Haykin \& Chen(2005)Haykin and Chen]{haykin2005cocktail}
Simon Haykin and Zhe Chen.
\newblock The cocktail party problem.
\newblock \emph{Neural computation}, 17\penalty0 (9):\penalty0 1875--1902,
  2005.

\bibitem[Hinton et~al.(2012)Hinton, Deng, Yu, Dahl, Mohamed, Jaitly, Senior,
  Vanhoucke, Nguyen, Sainath, et~al.]{hinton2012deep}
Geoffrey Hinton, Li~Deng, Dong Yu, George~E Dahl, Abdel-rahman Mohamed, Navdeep
  Jaitly, Andrew Senior, Vincent Vanhoucke, Patrick Nguyen, Tara~N Sainath,
  et~al.
\newblock Deep neural networks for acoustic modeling in speech recognition: The
  shared views of four research groups.
\newblock \emph{IEEE Signal Processing Magazine}, 29\penalty0 (6):\penalty0
  82--97, 2012.

\bibitem[Isola et~al.(2016)Isola, Zhu, Zhou, and Efros]{pix2pix2016}
Phillip Isola, Jun-Yan Zhu, Tinghui Zhou, and Alexei~A Efros.
\newblock Image-to-image translation with conditional adversarial networks.
\newblock \emph{arxiv}, 2016.

\bibitem[Larsen \& Aarts(2005)Larsen and Aarts]{larsen2005audio}
Erik Larsen and Ronald~M Aarts.
\newblock \emph{Audio bandwidth extension: application of psychoacoustics,
  signal processing and loudspeaker design}.
\newblock John Wiley \& Sons, 2005.

\bibitem[Ledig et~al.(2016)Ledig, Theis, Huszar, Caballero, Aitken, Tejani,
  Totz, Wang, and Shi]{DBLP:journals/corr/LedigTHCATTWS16}
Christian Ledig, Lucas Theis, Ferenc Huszar, Jose Caballero, Andrew~P. Aitken,
  Alykhan Tejani, Johannes Totz, Zehan Wang, and Wenzhe Shi.
\newblock Photo-realistic single image super-resolution using a generative
  adversarial network.
\newblock \emph{CoRR}, abs/1609.04802, 2016.
\newblock URL \url{http://arxiv.org/abs/1609.04802}.

\bibitem[Li et~al.(2015)Li, Huang, Xu, and Lee]{li2015dnn}
Kehuang Li, Zhen Huang, Yong Xu, and Chin-Hui Lee.
\newblock Dnn-based speech bandwidth expansion and its application to adding
  high-frequency missing features for automatic speech recognition of
  narrowband speech.
\newblock In \emph{Sixteenth Annual Conference of the International Speech
  Communication Association}, 2015.

\bibitem[Liang et~al.(2013)Liang, Hoffman, and
  Ellis]{DBLP:conf/ismir/LiangHE13}
Dawen Liang, Matthew~D. Hoffman, and Daniel P.~W. Ellis.
\newblock Beta process sparse nonnegative matrix factorization for music.
\newblock In Alceu de~Souza Britto~Jr., Fabien Gouyon, and Simon Dixon (eds.),
  \emph{Proceedings of the 14th International Society for Music Information
  Retrieval Conference, {ISMIR} 2013, Curitiba, Brazil, November 4-8, 2013},
  pp.\  375--380, 2013.
\newblock ISBN 978-0-615-90065-0.
\newblock URL
  \url{http://www.ppgia.pucpr.br/ismir2013/wp-content/uploads/2013/09/229_Paper.pdf}.

\bibitem[Liang et~al.(2015)Liang, Zhan, and Ellis]{liang2015content}
Dawen Liang, Minshu Zhan, and Daniel~PW Ellis.
\newblock Content-aware collaborative music recommendation using pre-trained
  neural networks.
\newblock In \emph{ISMIR}, pp.\  295--301, 2015.

\bibitem[Maas et~al.(2012)Maas, Le, ONeil, Vinyals, Nguyen, and Ng]{45168}
Andrew Maas, Quoc~V. Le, Tyler~M. ONeil, Oriol Vinyals, Patrick Nguyen, and
  Andrew~Y. Ng.
\newblock Recurrent neural networks for noise reduction in robust asr.
\newblock In \emph{INTERSPEECH}, 2012.

\bibitem[Mehri et~al.(2016)Mehri, Kumar, Gulrajani, Kumar, Jain, Sotelo,
  Courville, and Bengio]{mehri2016samplernn}
Soroush Mehri, Kundan Kumar, Ishaan Gulrajani, Rithesh Kumar, Shubham Jain,
  Jose Sotelo, Aaron Courville, and Yoshua Bengio.
\newblock Samplernn: An unconditional end-to-end neural audio generation model,
  2016.
\newblock URL \url{http://arxiv.org/abs/1612.07837}.
\newblock cite arxiv:1612.07837.

\bibitem[Odena et~al.(2016)Odena, Dumoulin, and Olah]{odena2016deconvolution}
Augustus Odena, Vincent Dumoulin, and Chris Olah.
\newblock Deconvolution and checkerboard artifacts.
\newblock \emph{Distill}, 2016.
\newblock \doi{10.23915/distill.00003}.
\newblock URL \url{http://distill.pub/2016/deconv-checkerboard}.

\bibitem[Park \& Kim(2000)Park and Kim]{park2000narrowband}
Kun-Youl Park and Hyung~Soon Kim.
\newblock Narrowband to wideband conversion of speech using gmm based
  transformation.
\newblock In \emph{Acoustics, Speech, and Signal Processing, 2000. ICASSP'00.
  Proceedings. 2000 IEEE International Conference on}, volume~3, pp.\
  1843--1846. IEEE, 2000.

\bibitem[Pulakka et~al.(2011)Pulakka, Remes, Palom{\"a}ki, Kurimo, and
  Alku]{pulakka2011speech}
Hannu Pulakka, Ulpu Remes, Kalle Palom{\"a}ki, Mikko Kurimo, and Paavo Alku.
\newblock Speech bandwidth extension using gaussian mixture model-based
  estimation of the highband mel spectrum.
\newblock In \emph{Acoustics, Speech and Signal Processing (ICASSP), 2011 IEEE
  International Conference on}, pp.\  5100--5103. IEEE, 2011.

\bibitem[Shi et~al.(2016)Shi, Caballero, Huszar, Totz, Aitken, Bishop,
  Rueckert, and Wang]{ShiCHTABRW16}
Wenzhe Shi, Jose Caballero, Ferenc Huszar, Johannes Totz, Andrew~P. Aitken, Rob
  Bishop, Daniel Rueckert, and Zehan Wang.
\newblock Real-time single image and video super-resolution using an efficient
  sub-pixel convolutional neural network.
\newblock pp.\  1874--1883, 2016.
\newblock \doi{10.1109/CVPR.2016.207}.
\newblock URL \url{http://dx.doi.org/10.1109/CVPR.2016.207}.

\bibitem[Sun \& Mazumder(2013)Sun and Mazumder]{DBLP:conf/mlsp/SunM13}
Dennis~L. Sun and Rahul Mazumder.
\newblock Non-negative matrix completion for bandwidth extension: {A} convex
  optimization approach.
\newblock In \emph{{IEEE} International Workshop on Machine Learning for Signal
  Processing, {MLSP} 2013, Southampton, United Kingdom, September 22-25, 2013},
  pp.\  1--6. {IEEE}, 2013.
\newblock \doi{10.1109/MLSP.2013.6661924}.
\newblock URL \url{http://dx.doi.org/10.1109/MLSP.2013.6661924}.

\bibitem[van~den Oord et~al.(2016)van~den Oord, Dieleman, Zen, Simonyan,
  Vinyals, Graves, Kalchbrenner, Senior, and
  Kavukcuoglu]{DBLP:journals/corr/OordDZSVGKSK16}
A{\"{a}}ron van~den Oord, Sander Dieleman, Heiga Zen, Karen Simonyan, Oriol
  Vinyals, Alex Graves, Nal Kalchbrenner, Andrew~W. Senior, and Koray
  Kavukcuoglu.
\newblock Wavenet: {A} generative model for raw audio.
\newblock \emph{CoRR}, abs/1609.03499, 2016.
\newblock URL \url{http://arxiv.org/abs/1609.03499}.

\bibitem[Wang \& Wang(2014)Wang and Wang]{wang2014improving}
Xinxi Wang and Ye~Wang.
\newblock Improving content-based and hybrid music recommendation using deep
  learning.
\newblock In \emph{Proceedings of the 22nd ACM international conference on
  Multimedia}, pp.\  627--636. ACM, 2014.

\bibitem[Yamagishi()]{yamagishienglish}
Junichi Yamagishi.
\newblock English multi-speaker corpus for cstr voice cloning toolkit, 2012.
\newblock \emph{URL http://homepages. inf. ed. ac.
  uk/jyamagis/page3/page58/page58. html}.

\bibitem[Zhang et~al.(2016)Zhang, Isola, and Efros]{zhang2016colorful}
Richard Zhang, Phillip Isola, and Alexei~A Efros.
\newblock Colorful image colorization.
\newblock \emph{ECCV}, 2016.

\end{thebibliography}
\bibliographystyle{iclr2017_workshop}

\end{document}